\documentclass[12pt]{article}
\usepackage[dvips]{graphicx}
\newcommand{\wav}{\mid\Psi>}
\begin{document}
\begin{flushright}
ISU-NP-THY-02-11
\end{flushright}
\vspace{0.2in}
\begin{center}
{Effective Operator Treatment of the Anharmonic Oscillator}
\vspace{.3in}

K.J. Abraham$^a$ \& J.P. Vary$^a$
\vspace{.3in}

{\small \it
$^a$ Department of Physics and Astronomy, Iowa State University,
     Ames, Iowa 50011, USA\\}
\end{center}
\vspace{.5in}
     \begin{center} ABSTRACT  \end{center}
We analyse the one-dimensional anharmonic oscillator using effective
operator  methods in both the strong and weak coupling limits. 
We show that in the
case of a one-dimensional model space, the  similarity transformation
needed to define the effective Hamiltonian is related to the
coefficients in the
expansion of the wave function in the unperturbed harmonic oscillator basis.
We obtain an
infinite system of equations which is equivalent to those obtained from the
Hill Determinant solution of the anharmonic  oscillator. 
The analytic properties of the resulting equations reveal the 
non-perturbative features of the  underlying problem. 
Thus, we demonstrate the
the utility of the effective operator method for solving a non-analytic strong
coupling problem.

\newpage
The effective operator method \cite{leesuz} has been used extensively and
successfully within a cluster approximation scheme 
to obtain the low-lying spectroscopy of complex nuclei
\cite{assorted} with realistic nucleon-nucleon interactions.
Central to the methodology of \cite{leesuz} is the iterative 
construction of a similarity transformation which transforms the original
hamiltonian to a new hamiltonian having a two-component block
diagonal structure where one component is finite
dimensional and accommodates the low-lying spectroscopy. 
Diagonalising this finite dimensional block diagonal sub-matrix 
yields a finite number of eigenvalues to any desired precision
corresponding to a subset of the exact solutions. In a sense,
the similarity transformation is designed to decouple a finite
dimensional sub-space from the rest of the spectrum,
even in cases when the original hamiltonian cannot be treated by perturbative 
methods. However, there are few rigorous results on the existence, 
or non-existence \cite{viazvar}, of the similarity transformation on which the 
utility of the method hinges, 
although several practical issues specific to strongly correlated nuclear
systems are under investigation. Thus the formal properties of this effective
operator approach require additional study, particularly within the 
framework of problems known to be non-perturbative in character. In addition
to formal aspects, a deeper understanding of the errors associated
with various approximations is required.
For example, when a sequence of clusters is introduced as in the nuclear
many body applications \cite{assorted} we still need to understand how to
optimize the convergence with increasing cluster size and/or with increasing
model space size. 
In order to clarify the physical utility of the 
approach and refine our knowledge of its limitations and properties we 
address a well studied non-perturbative problem, the one dimensional 
quartic oscillator in both the strong and weak coupling regime. In doing so,
we shed light on the origins of the decoupling which plays such a crucial role.

The hamiltonian $H$ that we select, has the form $H_{0} + V$ where
\begin{equation}
H_{0} = \frac{p^{2}}{2\,m} + \frac{m\omega^{2}x^{2}}{2}
\end{equation} and
$ V = \lambda\,x^{4} $.

The perturbation expansion for energy eigenvalues
is known not to converge, independent of the size of
$\lambda$
\cite{benderwu}. The
divergent behaviour of the expansion may be traced to the fact that, to
large orders in perturbation theory, the
growth in the number of contributions is sufficiently
rapid to lead to a series that eventually diverges, even for $\lambda$
arbitrarily small.

Various non-perturbative methods have been applied
to extract the eigenvalues \cite{general}. In this investigation
we will
construct an  effective Hamiltonian using a similarity transformation
along the lines proposed by Lee \& Suzuki \cite{leesuz}. In addition to 
extracting energy eigenvalues from the effective hamiltonian, 
the matrix elements of the generator of the similarity transformation will 
also be used to develop a system of equations equivalent to those
obtained from the Hill determinant solution of the anharmonic oscillator.

We begin by establishing the following notation: let $E$ denote a generic
eigenvalue and $\wav$ the corresponding
exact eigenstate of the Hamiltonian in Eq.1. In that case we have
\begin{equation}
H\wav = E\wav \label{eq:hamil} \end{equation}
Following \cite{leesuz}, we define 
$S$, the generator of a similarity
transformation such that $\tilde{H} = e^{-S}\,H\,e^{S}$ and
$\tilde{\wav} = e^{-S}\wav$ . It then follows that \begin{equation}
\tilde{H}\tilde{\wav} = E\tilde{\wav} \end{equation}
For now, $S$ is arbitrary but subsequently, restrictions will be placed on $S$.

The next step in the treatment along the lines of \cite{leesuz} is the
identification of a suitable model space (P-Space). We will use a one
dimensional model space containing the ground state of $H_{0}$ signified by
$\mid 0>$. With this choice, P, the projection operator onto the model
space is just
$\mid 0><0 \mid$. The Q space is then $\Sigma\mid i><i\mid$, where the sum
over $i$ runs over all states of the harmonic oscillator other
than the ground state; $P$ and $Q$ are orthogonal and $P + Q = I$.

The first constraint we impose on $S$ is that 
$S\,=\,QSP$. This leads to the identities
\begin{displaymath}
PS = SQ = 0 \end{displaymath} and 
\begin{displaymath}
S = SP = QS 
\end{displaymath} 
These identities will be used extensively
later on. Since $P$ and $Q$ are orthogonal, $S ^{2}= 0$ leading to
$e^{\pm S} = I \pm S$. 
With this choice of
$P$ and $Q$ spaces and the restriction on $S$, 
$S$ reduces to a column vector whose non-vanishing
matrix elements are of the form $<i\mid S\mid 0>\,\,(i\neq 0)$ which
will be denoted by  $S_{i}$. The $S_{i}$ will be seen to be related to the 
expansion coefficients of the true wavefunction in the unperturbed basis.
The final condition we impose on $S$ comes from requiring that \begin{equation}
\tilde{H}P\tilde{\wav} = E P \tilde{\wav}\label{eq:hamils}\end{equation} 

As in \cite{leesuz} we define the effective hamiltonian $H_{eff}$ to be 
$P\tilde{H}P$. In general, diagonalising $H_{eff}$ yields a subset of
energy eigenvalues of the full hamiltonian, since in our case
$H_{eff}$ is one dimensional we will recover just one eigenvalue for any 
given $S$. 
Expanding $H_{eff}$ in terms of $S$ gives 
\begin{equation} H_{eff} = PHP + PVS \label{eq:heff} \end{equation}
From Eq.~\ref{eq:hamils} and the fact that $P$ and 
$Q$ are projection operators it follows that $ Q\tilde{H}P = 0 $, yielding 
\begin{equation}
SHP + SVS = QVP + QHS \label{eq:opeqn} \end{equation} which can be 
rewritten as
\begin{displaymath}
QVP + QHS = SH_{eff}
\end{displaymath}
 
After  subtracting $\Omega\,S$ from both sides of the preceeding equation, 
($\Omega$ is an arbitrary parameter),
we obtain the following equation for $S$ \begin{equation}
S = A(QVP -S(H_{eff} -\Omega)) \label{eq:sdef}
\end{equation}
where \begin{displaymath}
A = {(\Omega -QHQ)}^{-1}
\end{displaymath} 
Note that since $S$ and $H_{eff}$ must be $\Omega$ independent, 
$\Omega$ is strictly speaking redundant. Nonetheless, the utility of 
$\Omega$ will soon be made clear. 
No use has been made of the fact that the P space is one 
dimensional, thus Eqs.~\ref{eq:hamils}, ~\ref{eq:heff} \& \ref{eq:sdef} are perfectly general 
and identical equations arise when treating more complex hamiltonians along 
the lines of \cite{leesuz}.

Since $H_{eff}$ depends explicitly on $S$, Eq.~\ref{eq:sdef} is a non-linear
operator equation for $S$ potentially admitting more than one solution. In 
addition, $A$ is ill-defined in an infinite dimensional Hilbert Space.
In applications to many body problems it is customary to impose an arbitrary 
cut-off in the Q space which leads to a well defined $A$, in our case
this amounts to neglecting all oscillator quanta above a certain energy. This
also reduces $S$ to a finite dimensional column vector. Even with this 
simplification Eq.~\ref{eq:sdef} is still hard to solve exactly; in 
applications to many body hamiltonians iterative solutions are preferred. We 
propose the following sequence of iterations; 
\begin{eqnarray} 
S^{(n)} & = & A(QVP -S^{(n-1)}({H^{(n)}}_{eff} -\Omega)) \\
{{H}^{(n+1)}}_{eff} & = & PHP + PVS^{(n)} 
\end{eqnarray} 
with $S^{(0)} = 0$ as the starting point in the sequence. After a suitable 
number of iterations the sequence is terminated. 
Since the $P$ space 
is one dimensional, the final $H_{eff}$ is simply the energy eigenvalue, no 
diagonalisation is necessary. On the other hand, it is necessary
to check the cut-off independence of the energy eigenvalues and the 
convergence of the iterations, this can be done by varying both the 
cut-off and $\Omega$. As a check we calculate the eigenvalues emerging from 
this procedure after a total of seven iterations.
We retain only the first 
ten harmonic oscillator eigenstates of even parity in calculating $A$; since
$V$ does not mix eigenstates of even and odd parity it is adequate to 
consider only even parity states in calculating $A$.
As in
\cite{biswas} we use $\hbar = 1$, $\omega = 2$ and $m = .5$ and allow
$\lambda$ to vary. Our results (setting $\Omega = -1$) are 
\newline
\vspace{.3cm}
\newline
\begin{tabular}{|l|l|l|l|l|l|l|l|} \hline
Iteration$(n)$ & $\lambda = .1$ &$\lambda = .2$ &$\lambda =.3$& $\lambda =1$ & 
 $\lambda = 2$ & $\lambda = 3$ \\ \hline
0& 1.075   & 1.15    & 1.225   & 1.75    & 2.5     & 3.25     \\ \hline
1& 1.06792 & 1.12612 & 1.17806 & 1.45479 & 1.7359  & 1.95799  \\ \hline
2& 1.06603 & 1.12035 & 1.16752 & 1.40423 & 1.62662 & 1.79234  \\ \hline
3& 1.06551 & 1.11886 & 1.16495 & 1.39484 & 1.61088 & 1.77311  \\ \hline
4& 1.06535 & 1.11845 & 1.16429 & 1.3929  & 1.60819 & 1.77022  \\ \hline
5& 1.06531 & 1.11834 & 1.16411 & 1.39248 & 1.60768 & 1.76971  \\ \hline 
6& 1.06529 & 1.11831 & 1.16407 & 1.39238 & 1.60757 & 1.76962  \\ \hline 
\end{tabular}
\newline
\vspace{.3cm}
\newline
These results are in excellent agreement with the ground state energies 
in \cite{biswas}; 
extending the size 
of the $Q$ space used to obtain $A$ changes the eigenvalues (if at all)
only at the fifth significant figure or beyond. Furthermore, 
allowing $\Omega$ to vary between -2 to +2 leads to a similarly small change in 
the obtained eigenvalues. As in \cite{varbarr} we can access different 
solutions for $S$ and thus different eigenvalues, by choosing a different range
of values for $\Omega$.  

Given that the hamiltonian is non-perturbative the 
convergence of the sequence of iterations is non-perturbative in character;
furthermore the error induced by truncating the $Q$ space to 
construct $A$ is small. 
Since truncation of the $Q$ space in the manner just described is a key 
ingredient in the application of effective operator methods in many 
body hamiltonians, we will attempt to understand why this procedure is valid.
More precisely, we will show that at least for the purposes of 
calculating low lying eigenvalues, $S$ may be accurately 
approximated by a column vector of finite dimension; since $S = QSP$, it 
then follows the $Q$ space needed to define $A$ is effectively finite 
dimensional which justifies the truncation we employ.

We start by expanding Eq.~\ref{eq:hamils} in terms of $S$ and $H$ leading to
\begin{equation}
(H -SH + HS - SHS)P\wav = E P \wav \label{eq:master} \end{equation}

In order to derive an identity for $E$ in terms of the $S_{i}$ we expand
$\wav$ in terms of the eigenstates of $H_{0}$ as follows
\begin{equation} \wav = \sum \alpha_{n}\mid n> \label{eq:expand}
\end{equation} The sum in Eq.~\ref{eq:expand} begins from zero and runs over
all positive integers.
For the sake of future notational convenience we will
denote $ <n\,\mid A\,\mid m>$  by $A_{nm}$ where $A$ is any operator.
Inserting the form of $\wav$  from Eq.~\ref{eq:expand}
into Eq.~\ref{eq:master} and operating with $<\,0\mid$ on both sides
yields \begin{equation}
E = <\,0\mid H \mid\,0>  + <\,0\mid HS \mid\,0> \end{equation} assuming
that $\alpha_{0}$ is non-vanishing.
Inserting a complete set of states yields
\begin{equation}
E = H_{00} +
V_{02}S_{2} +
V_{04}S_{4} \label{eq:engeq} \end{equation}
exploiting the fact that the quartic perturbation connects the ground state
only to states $\mid 2>$ and $\mid 4>$. Eq.~\ref{eq:engeq} is a
special case of a more general expression for the
exact eigenenergy resulting from an effective hamiltonian in a one-dimensional
model space derived in \cite{varbarr}.

Equations for the remaining $S_{i}$ may be obtained by sandwiching 
Eq.\ref{eq:opeqn} between $\mid 0>$ and $<n\mid$ where $n\neq 0$. Doing so and
using the fact that $P$ and $Q$ are projection operators gives \begin{equation}
<n\mid SH\mid 0> + <n\mid SVS\mid 0> \,
= \, <n\mid V\mid 0>  + <n\mid HS \mid \,0>
\label{eq:matelem} \end{equation}

By judicious insertions of complete sets of states, the left hand side of
Eq.~\ref{eq:matelem} can be reduced to
\begin{displaymath}
S_{n}(H_{00} + V_{02}S_{2}  +
V_{04}S_{4} ) \end{displaymath} The term within brackets
is just $E$ from Eq.~\ref{eq:engeq}. Thus Eq.~\ref{eq:matelem} may be
written as
\begin{equation}
S_{n}E = <n\mid V\mid 0>  + <n\mid HS \mid 0>
\end{equation}
Inserting a complete set of states yields
\begin{equation}
S_{n}E = V_{n0}  + \sum_{m} H_{nm}S_{m}
\label{eq:hilldet}
\end{equation}
As before we will  assume $n$ to be even, the extension to include odd $n$ 
is straightforward.
Inserting different values of $n$ into Eq.~\ref{eq:hilldet} leads to
a system of coupled equations for the $S_{n}$.
For $n=2$ we have,
\begin{equation}
V_{20} + (H_{22} -E)S_2 + V_{24}S_{4} +V_{26}S_{6} = 0
\label{eq:hilln2}
\end{equation}
For $n=4$ we have,
\begin{equation}
V_{40} + V_{42}S_{2} + (H_{44} -E)S_4 + V_{46}S_{6} + V_{48}S_{8} = 0
\label{eq:hilln4}
\end{equation}
For $n=i$ and $i>4$, there is no term independent of $S$. The equation takes
the form
\begin{equation}
V_{ii-4}S_{i-4} + V_{ii-2}S_{i-2} + (H_{ii} -E)S_{i} + V_{ii+2}S_{i+2}
+V_{ii+4}S_{i+4} = 0
\label{eq:hillni}
\end{equation}

\noindent We can make all the preceeding equations homogenous with the generic
substitution
\begin{displaymath}
S_{i} = \frac{\beta_{i}}{\beta_{0}}
\end{displaymath}
With this substitution Equations~\ref{eq:engeq}, \ref{eq:hilln2},
\ref{eq:hilln4} read
\begin{eqnarray}
(H_{00} - E)\beta_{0} + V_{02}\beta_{2} + V_{04}\beta_{4} & = & 0 \\
V_{20}\beta_{0} + (H_{22} - E)\beta_{2} + V_{24}\beta_{4} + V_{26}\beta_{6}&
= & 0 \\
V_{40}\beta_{0} + V_{42}\beta_{2} + (H_{44} - E)\beta_{4} + V_{46} \beta_{6} +
V_{48}\beta_{8}&
= & 0 
\end{eqnarray}
For larger values of $n$ it is adequate to replace $S_{i}$ by $\beta_{i}$ in
Eq.~\ref{eq:hillni} giving \begin{equation}
V_{ii-4}\beta_{i-4} + V_{ii-2}\beta_{i-2} + (H_{ii} -E)\beta_{i} +
V_{ii+2}\beta_{i+2}
+V_{ii+4}\beta_{i+4} = 0
\end{equation}

\noindent For the sake of comparison, we substitute the expansion from
Eq.~\ref{eq:expand}
into Eq.~\ref{eq:hamil}. Restricting ourselves to $\alpha_{i}$ with $i$ even
yields
\begin{eqnarray}
(H_{00} - E)\alpha_{0} + V_{02}\alpha_{2} + V_{04}\alpha_{4} & = & 0 \\
V_{20}\alpha_{0} + (H_{22} - E)\alpha_{2} + V_{24}\alpha_{4} +
V_{26}\alpha_{6}&
= & 0 \\
V_{40}\alpha_{0} + V_{42}\alpha_{2} + (H_{44} - E)\alpha_{4} +
V_{46}\alpha_{6} +
V_{48}\alpha_{8}&
= & 0 
\end{eqnarray} and
for larger values of $i$ we have \begin{equation}
V_{ii-4}\alpha_{i-4} + V_{ii-2}\alpha_{i-2} + (H_{ii} -E)\alpha_{i} +
V_{ii+2}\alpha_{i+2}
+V_{ii+4}\alpha_{i+4} = 0
\end{equation}

\noindent It is clear that $\alpha_{i}$ and $\beta_{i}$ satisfy the same
set of
equations. However, solving for the $\alpha_{i}$ is equivalent
to solving the Hill determinant for the quartic oscillator \cite{biswas}.
Since the $S_{i}$ are proportional to the $\beta_{i}$, the $S_{i}$ may also be 
obtained from solving the Hill determinant.
Furthermore, the physical significance of the
$S_{i}$ is apparent, the $S_{i}$ may be taken equal (up to an overall
constant) to the
coefficients arising in the expansion of the true wave function in the
unperturbed basis.
As far as we are aware, this is the first instance where a connection between 
$S$ and the expansion of the wave function in unperturbed basis has been 
established. 

The $\alpha_{i}$ may be independently determined by the requirement that
\begin{displaymath}
det(H -EI) = 0
\end{displaymath} 
for an invertible $H$. 
It can be shown that this requirement is equivalent to the more general
statement
in \cite{varbarr} of which Eq.~\ref{eq:engeq} is a special case
\cite{sprung}. This
provides an independent cross-check of our results.

There are additional constraints on the $S_{i}$ for large $i$ which arise
from the behaviour
of the matrix elements $V_{ij}$ at large $i$. To derive these constraints,
let us rewrite
Eq.~\ref{eq:hillni} as follows, \begin{equation}
V_{ii-4}S_{i-4} + V_{ii-2}S_{i-2} + H_{ii}S_{i} + V_{ii+2}S_{i+2}
+V_{ii+4}S_{i+4} = ES_{i}
\label{eq:lasts}
\end{equation}
For large $i$ the matrix elements appearing in Eq.~\ref{eq:lasts} have the
approximate form
\begin{eqnarray}
<i\mid x^{4}\mid (i+4) > & = &
(\frac{\hbar}{m\omega})^{2}\,(\frac{1}{4})\,i^{2}(1 + 5/i) + \cdots  \\
<i\mid x^{4}\mid (i+2) > & = &
(\frac{\hbar}{m\omega})^{2}\,i^{2}\,(1 + 3/i) + \cdots  \\
<i\mid x^{4}\mid i > & = &
(\frac{\hbar}{m\omega})^{2}\,(\frac{3}{4})\,(2i^{2} +2i + 1) \\
<i\mid x^{4}\mid (i -2) > & = &
(\frac{\hbar}{m\omega})^{2}\,i^{2}\,(1 - 1/i) + \cdots  \\
<i\mid x^{4}\mid (i-4) > & = &
(\frac{\hbar}{m\omega})^{2}\,(\frac{1}{4})\,i^{2}(1 - 5/i) + \cdots  \\
<i\mid x^{2}\mid (i+2) > & = &
(\frac{\hbar}{2\,m\omega})\,i +\cdots \\
<i\mid x^{2}\mid (i-2) > & = &
(\frac{\hbar}{2\,m\omega})\,i + \cdots \\
<i\mid x^{2}\mid i > & = & (\frac{\hbar}{2\,m\omega})(2i +1)
\end{eqnarray}
where higher order finite terms in the expansion have been neglected.
As can be seen from the above equations, the left hand side of
Eq.~\ref{eq:lasts}
contains both quadratic and linear 
divergences in $i$ at large $i$,
and no such divergences appear on the right.
Eq.~\ref{eq:lasts} will be consistent only if
$S_{i}$ fall off rapidly for large $i$ aided by possible cancellations due
to sign
differences between the various $S_{i}$ appearing on the left. 
As a
consequence of the $S_{i}$ falling off rapidly,
low energy eigenvalues of the full hamiltonian are expected to be 
only weakly 
dependent on high energy eigenstates of the  unperturbed hamiltonian.
As a check, we allow $i$ in Eq.~\ref{eq:lasts} to run over a limited 
range of even values beginning at 0,
and compute the lowest energy eigenvalues from the resulting 
equations. 
We use the same numerical values for the parameters as before and allow 
$\lambda$ to vary over the same range.
The results are summarised in the table below:
\newline
\vspace{.5cm}
\newline
\begin{tabular}{|l|l|l|l|l|l|l|l|} \hline
$imax/2\,\downarrow$ & $\lambda\,\rightarrow$   &.1    &.2    &.3    &1
&2 & 3\\ \hline
5 & &1.06529  & 1.11829  & 1.16406  &1.39337  &1.61123 &1.77481  \\ \hline
10& & 1.06529 & 1.11829  & 1.16405  & 1.39235 &1.60755  &1.76963 \\ \hline
15& & 1.06529 & 1.11829  & 1.16405  & 1.39235 & 1.60754 & 1.76959 \\ \hline
20&  & 1.06529 & 1.11829  & 1.16405  & 1.39235 & 1.60754 & 1.76959 \\ \hline
\end{tabular}
\newline
\vspace{.5cm}
\newline
The entries in the table are the lowest energy eigenvalues obtained from
sets of not more than
20 equations, and are in excellent agreement with the results obtained by 
our iterative procedure and with \cite{biswas}. This is strong
support of our earlier claim that low energy eigenvalues of the full
hamiltonian are only weakly 
dependent on high energy states of the unperturbed hamiltonian.
It is now clear why the truncation of the $Q$ space which was implemented 
in order to facilitate the iterative procedure is justified.
The decoupling of high energy states 
in a toy model has been 
independently studied in \cite{bog}; however unlike the quartic oscillator the 
hamiltonian in \cite{bog} can be analytically inverted, permitting the use of 
methods very different from the ones we employ in this paper.

As an additional analytical check on the decoupling of $S_{i}$ for large $i$,
we consider the following Ansatz for large $i$, \begin{equation}
\frac{S_{i+j}}{S_{i}} \sim
(-1)^{j/2}(1 - \frac{jf(j/2)}{i}\frac{m^{2}\omega^{3}}{\lambda\hbar})
\label{eq:srecur}
\end{equation}
where $f$ is some arbitrary unspecified function. Inserting this Ansatz into
Eq.~\ref{eq:lasts} and using the asymptotic forms of the matrix elements given
earlier, we see that
Eq.~\ref{eq:lasts} is free of
divergences in $i$
provided \begin{equation}
f(-2) -f(2) + 2f(1) -2f(-1) + 1 = 0
\label{eq:fcond} \end{equation}
The utility of Eqs.~\ref{eq:fcond} \& ~\ref{eq:srecur} will become apparent
when
we investigate what happens when $\lambda$ is so large that it
dominates the quadratic
term in the potential. In that case the hamiltonian may be conveniently
expressed as
\begin{displaymath}
\frac{p^{2}}{2\,m} + \frac{m\omega^{2}x^{2}}{2}
+ \lambda x^{4} - \frac{m\omega^{2}x^{2}}{2}
\end{displaymath}

Following the same procedure as before but with
$ V= \lambda x^{4} - \frac{m\omega^{2}x^{2}}{2}$ gives an equation
identical to
Eq.~\ref{eq:lasts} but with different values of $H_{ii}$, and $V_{ii\pm
2}$. This is
not surprising as the structure of Eq.~\ref{eq:lasts} does not rely on the
precise
form of $V$ but on the fact that the only non-vanishing matrix
elements
$V_{ij}$ have
$\mid(i-j)\mid \leq 4$, which is the case for both forms of $V$ that we
consider.

Once again,
consistency requires that all divergences quadratic and linear in $i$
cancel on the left
hand side of Eq.~\ref{eq:lasts} with $V$ modified to study the strong
coupling limit. Using the approximate forms of the relevant matrix elements
given
earlier, it is straightforward to verify that the recursion relation in
Eq.~\ref{eq:srecur} originally
derived to analyse finite coupling is sufficient to gaurentee the
cancellations in the
strong coupling limit as well. The only difference between the two cases lies
presumably only in
the ${\cal O}(1/i^{2})$ term which plays no role in cancellation of
divergences. This is a further indication of the non-perturbative
nature of the effective operator method. 

To conclude, we have implemented an 
an effective operator treatment of the anharmonic oscillator on the lines 
of \cite{leesuz}. We have analyzed the validity of a key simplification made 
in the treatment of the nuclear many body problem. The role of the expansion 
of the exact
eigenfunction in the basis of the unperturbed hamiltonian in defining
$S$, the generator of the similarity transformation has been 
emphasized. Both the strong and weak coupling cases may be treated along the
same lines, underscoring the non-perturbative nature of the formulation. Our
numerical application demonstrates rapid convergence to known results with
modest effort.

This work was supported in part by U.S. DOE Grant No. DE-FG-02-87ER-40371,
Division of High Energy and Nuclear Physics.

\end{document}